\documentclass[showpacs,amsmath,amssymb]{revtex4}

\usepackage{graphicx}% Include figure files
\usepackage{dcolumn}% Align table columns on decimal point
\usepackage{bm}% bold math

\usepackage[ansinew,latin1]{inputenc}
\usepackage{psfrag}
\usepackage{array}

\usepackage{physique}

\begin{document}

\title{Nanoparticles as a possible moderator for \\ an ultracold neutron source}

\author{V.V. Nesvizhevsky}
	\email{nesvizhevsky@ill.fr}
\author{G. Pignol}%
 	\email{guillaume.pignol@polytechnique.org}
 \affiliation{Institute Laue Langevin, Grenoble, France}
\author{K.V. Protasov}
	\email{protasov@lpsc.in2p3.fr}
 \affiliation{Laboratoire de Physique Subatomique et de Cosmologie,
 CNRS/IN2P3-UJF, 53, Avenue des Martyrs, Grenoble, France}

\date{\today}% It is always \today, today,

\begin{abstract}
Ultracold and very cold neutrons (UCN and VCN) interact strongly
with nanoparticles due to the similarity of their
wavelengths and nanoparticles sizes. We analyze the hypothesis that
this interaction can provide efficient cooling of neutrons by
ultracold nanoparticles at certain experimental conditions, thus
increasing the density of UCN by many orders of magnitude. The
present analytical and numerical description of the problem is
limited to the model of independent nanoparticles at zero
temperature. Constraints of application of this model are
discussed.
\end{abstract}


\maketitle

%--------------------%
\section{Introduction}
%--------------------%

A series of experiments performed by a number of research groups
have brought light to the phenomenon of the quasi-elastic
scattering of UCN at surfaces displaying surprisingly small energy
changes in the order of $10^{-7}$ eV \cite{losstrap1, losstrap2,
losstrap3, losstrap4, losstrap5, losstrap6, losstrap7}. A detailed
study of this process \cite{surfacenanoparticles1,
surfacenanoparticles2, surfacenanoparticles3,
surfacenanoparticles4, surfacenanoparticles5,
surfacenanoparticles6} has allowed us to conclude that, for solid
surfaces at least, this is due to the inelastic coherent
scattering of UCN on nanoparticles or nanostructures weakly
attached to the surface in a state of permanent thermal motion.
This conclusion triggered the idea considered in this article of neutron cooling at
ultracold nanoparticles. Complete control of the corresponding UCN losses from storage
bottles and/or a significant increase in UCN density are of
utmost importance for neutron-based research in fundamental
physics. This research includes the measurement of the neutron lifetime
\cite{lifetime1, lifetime2, lifetime3, lifetime4, lifetime5}, the
search for the non-zero neutron electric dipole moment
\cite{nEDM1, nEDM2}, the study of the gravitationally bound
quantum states of neutrons \cite{gravitation1, gravitation2,
gravitation3}, and the search for the non-zero neutron electric
charge \cite{charge}.

In any experiment with trapped neutrons, nanoparticle
temperature is equal to the trap temperature $T$ in the typical
range of $10$-$10^3$ K, while UCN energy corresponds to UCN
temperature T $\approx$1 mK. Ultracold neutrons therefore
preferentially increase their energy in collisions with such
``warm" nanoparticles. The probability of such inelastic UCN
scattering on the surface is small, since the surface density of
such weakly attached nanoparticles is typically small. 

However, the problem of neutron heating due to neutron-nanoparticle
interaction can in principle be reversed: the interaction of warm
neutrons with ultracold nanoparticles at a temperature of
$\approx$1 mK can cool down the neutrons
\cite{surfacenanoparticles1, GP, VVN2}. If the density of weakly
bound nanoparticles is high (these nanoparticles not only cover
the surface but also fill the volume) and if, as the neutrons
cool, the probability of their absorption and $\beta$-decay is
low, the neutron density will increase. This process can, for the
first time, allow equilibrium cooling of neutrons down to the UCN
temperature. One should note that in this case the moderator temperature should be as low as the UCN temperature in contrast to traditional methods to produce UCN.

The cooling of neutrons in nuclear reactors and spallation sources by a factor 
of about 10$^8$! is achieved by just a few
dozen collisions with nuclei in reactor moderators (hydrogen,
deuterium). The energy transfer is very efficient since the mass
of the moderator nuclei is equal to (or approximates) the neutron
mass and the neutron losses during the cooling process are low due
to small number of collisions needed to slow down neutrons.
However, no further efficient cooling occurs: the lower the neutron energy,
the larger the neutron wavelength. When the wavelength becomes
commensurate with the distance between the nuclei of the
moderator, the neutrons do not ``see" individual nuclei any longer
-- they are just affected by the average optical potential of the
medium. 
A further cooling of the neutrons due
to their interaction with collective degrees of freedom (such as
phonons) is less efficient than the moderation of the neutrons due
to their collisions with nuclei. That is insufficient however to cool the
main portion of the neutrons to the UCN energy region \cite{othersources1, othersources2, othersources3, othersources4,
othersources5, othersources6}. 

The idea of neutron cooling on
ultracold nanoparticles consists in reproducing the principle of
neutron cooling in reactor moderators using multiple collisions.
However, there is a difference in scale: the sizes of scattering centers 
are greater by a factor
of $\approx 10^2$; this increases the energy range of application
of this mechanism by a few orders of magnitude. 

It should be noted
that a UCN source of this type is based on the principle of UCN density
accumulation, as in a super-thermal source \cite{othersources3},
but not on the use of a UCN flux from a source in the flow-through
mode. In conventional sources used to select UCN, thermal
equilibrium is not achieved. These sources are much hotter than
UCN. Only a very small portion of the neutrons is used -- the other
neutrons are lost. Actually, these are sources of cold or very
cold neutrons (VCN), and experimentalists have to select a narrow
fraction of a broad energy spectrum. For instance, the most
intense flux of UCN available for users is now produced in a
liquid-deuterium source placed within the core of the high-flux
reactor at the Institut Laue-Langevin (ILL) \cite{othersources2}.
It increases the UCN flux by a factor of about $10^2$ in relation
to that available otherwise in the reactor in the thermal
equilibrium spectrum. Only a fraction of the neutron flux of about
$10^{-9}$ is thus actually used. On the other hand, the cooling of
neutrons on ultracold nanoparticles could provide for further
neutron cooling in a significant energy range, thereby increasing
the neutron density.

The new method for producing UCN consists in the equilibrium
cooling of VCN -- through their many collisions with ultracold
nanoparticles made from low-absorption materials (D$_2$, D$_2$O,
O$_2$ etc.) -- down to the temperature of these nanoparticles of
$\approx 1$ mK, during the diffusion motion of these neutrons in a
macroscopically large body of nanoparticles. The principle of
equilibrium cooling allows an increase in the neutron phase-space
density, in contrast to the method of selecting a narrow energy
range out of a warmer neutron spectrum. The use of nanoparticles
provides a sufficiently large cross section for coherent
interaction and an inhomogeneity of the moderator density, on a
spatial scale of about the neutron wavelength; it also shifts the
energy transfer range far below a value of about $10^{-3}$ eV, the
characteristic limit for liquid and solid moderators. Many
collisions are needed since the mass of the nanoparticles is much larger
than the neutron mass; the energy transfer to nanoparticles and
nanostructures is only moderately efficient. The need for a large
number of collisions limits the choice of materials: only low
absorption materials are appropriate. The temperature of the nanoparticles
 must correspond to the minimal energy to which
neutrons can still be cooled using this method. The diffusion
motion of neutrons in the body of nanoparticles allows us to
minimize the thermalization length and, accordingly, to increase
the achievable UCN density.

The cooling itself is provided by the interaction of neutrons with
individual degrees of freedom of weakly bound or free
nanoparticles, as well as by the excitation of collective degrees
of freedom in the body of nanoparticles (e.g. vibrations and
rotations), and also by the breaking of inter-particle bonds.
Details about gels of nanoparticles can be found in \cite{Gel1,
Gel2, Gel3, Gel4}. Even free nanoparticles in the gel have several
degrees of freedom: rotation and translation. In this paper, we
provide detailed calculations of the cooling of neutrons in a gel of
nanoparticles, considering only the collisions on free
nanoparticles, and neglecting the rotation. The interaction between the nanoparticles, including the long-range interaction induced by helium, is neglected here. So we deal with an
idealized gas of free nanoparticles at $0 \text{ K}$ in superfluid
helium. Our goal is to determine the behavior of a neutron in such
a moderator. 

In section \ref{model}, we give a model for the
interaction of a neutron with a single nanoparticle in suspension
in liquid helium. This model leads to a complete quantum solution
for three main quantities, which are, the absorption cross section
$\sigma_a$, the total scattering cross section $\sigma_s$, and at
last, the mean relative energy loss per collision $\xi =
\frac{\left\langle \Delta E\right\rangle}{E}$. This allows us to
describe the slowing down of neutrons in a gas of free
nanoparticles. In section \ref{infmoderator}, we deal with the
ideal situation of the infinite moderator, an infinite medium made
with nanoparticles surrounded by helium at $0$ K, and where the
only loss of neutrons are due to the absorption by a nanoparticle
($\beta$-decay is neglected). This naive model provides necessary
conditions for an efficient moderation. In section
\ref{realmoderator}, we will present estimations for finite
moderators, the characteristic size and the characteristic time of
thermalization. We compare efficiency of the moderator with
nanoparticles made of different materials, and also for different
nanoparticle sizes. We first chose deuterium as the material
to illustrate the calculations.

%------------------------------------------------%
\section{\label{model}Free nanoparticles model}
%------------------------------------------------%

Let us consider a nanoparticle with radius R, made of hundreds of
nuclei, immersed into superfluid helium. A low energy incident
neutron only sees the average potential of each nanoparticle. So we
will assume the following phenomenological potential for the
interaction of a neutron with the considered nanoparticle:
\begin{eqnarray}\label{pot1}
V(\vect{r}) = \left\{
\begin{array}{ll}
V \equiv V_0 - i V_1 & \text{  if  } r < R,  \\
0           & \text{  if  } r > R.
\end{array} \right.
\end{eqnarray}
We will provide detailed calculations of the collision parameters
in the Born approximation using this model. But let us first
estimate all the parameters of the potential (\ref{pot1})
describing neutron-nanoparticle interaction.

\subsection*{Parameters of neutron-nanoparticle interaction potential}

The potential $V_0 - i V_1$ is taken to be the averaged of each
nucleus {\sl Fermi} potential in the nanoparticle (neutron-helium interaction potential has to be subtracted):
\begin{equation}
V_0 - i V_1 = \sum_j \rho^{(j)} \int V^{(j)}(\vect{r}) d \vect{r} - V_{\text{He}}
\end{equation}
where the sum is done over the different kinds of nuclei in the
nanoparticle -- for instance, for a heavy water
nanoparticle, we must take into account the contributions of both
deuterium and oxygen nuclei. In each term of the sum, $\rho^{(j)}$ is the number of nuclei
of type $j$ per volume unit, and $V^{(j)} = V^{(j)}_0 - i
V^{(j)}_1$ is the interaction potential between a neutron and a
$j$-type nucleus. The real part of this potential can be derived
from the coherent scattering length $b^{(j)}$
\begin{equation}
b^{(j)} = \frac{m}{2 \pi \hbar^2} \int V^{(j)}_0 d \vect{r}
\end{equation}
where $m$ is the neutron mass. The imaginary part of the
potential, which describes the possibility of the neutron
capture, can be calculated from the absorption cross section via
the optical theorem, taking into account the fact that this
interaction is not strong and can be treated within the first Born
approximation:
\begin{equation}
\frac{2 m}{\hbar^2} \int V_1^{(j)} d \vect{r} = \sigma_a^{(j)}(k) k.
\end{equation}
The experimental data are given for thermal neutrons, i.e. for a
neutron velocity of $2200$ m/s. So we express the result in term
of $\sigma_a^{(j)}(k_0) k_0$, where $k_0$ is the wave vector of a
thermal neutron. Eventually:
\begin{eqnarray}
V_0 & = & 2 \pi \frac{\hbar^2}{m} \sum_j \rho^{(j)} b^{(j)},\\
V_1 & = & \frac{\hbar^2}{2 m} \sum_j \rho^{(j)} \sigma_a^{(j)}(k_0) k_0.
\end{eqnarray}

\begin{table}
\begin{center}
\caption{\label{nuclear}Nuclear data used in these calculations.
The experimental data are taken from \cite{NIST}.}
\begin{ruledtabular}
\begin{tabular}{cccc}
Nucleus  &  $b$ (fm)  &  $\sigma_a$ (mbarn) & Mass (m)\\
\hline
$^{1} $H   &  $-$3.74  &  333                         & 0.999  \\
$^{2} $H   &   6.67  &  0.52                        & 1.997  \\
$^{3} $He  &         &  5.33$\cdot 10^6$  & 3.968  \\
$^{4} $He  &   3.26  &  0                               & 3.968  \\
       Be  &   7.79  &  7.6                         & 8.935  \\
       C   &   6.65  &  3.5                         & 11.91  \\
       O   &   5.80  &  0.19                        & 15.86  \\
\end{tabular}
\end{ruledtabular}
\end{center}
\end{table}

From the nuclear data shown in table \ref{nuclear}, we can
compute all properties of the  potential of interaction between
neutrons and nanoparticles in superfluid helium. The results are
shown in table \ref{nanoparticle}, for various materials, taking
into account the following points:

\begin{itemize}
\item We note $A$ the nanoparticle mass in units of neutron mass.
Its variation with the radius is determined by the parameter
$A_0$: $A = A_0 R^3$.

\item The density of liquid helium is $124.9$ kg m$^{-3}$ at
boiling point $4.25$ K. We calculate the effective potential of the
nanoparticles in liquid helium which has a potential
$V_{\text{He}} = 15.9$ neV.

\item We calculate the density of nanoparticles (number of
nanoparticles in a unit volume) $N = N_0 R^{-3}$, assuming that
the total mass of the nanoparticles is $1\%$ of the total mass of
the helium. This parameter is only useful for the description of a
finite moderator in section \ref{realmoderator}.
\end{itemize}

\begin{table}
\begin{center}
\caption{\label{nanoparticle}Nanoparticle characteristics.}
\begin{ruledtabular}
\begin{tabular}{cccccc}
Nanoparticle & Density                      &  $A_0$             &  $N_0$  &  $V_0$  &  $V_1$ \\
             & ($\text{kg} / \text{m}^{3}$) & ($\text{nm}^{-3}$) &         &  (neV)  &  (neV) \\
\hline
D$_2$        & 195  & 488  & 0.001510 & 85 & 2.2 $\cdot 10^{-6}$ \\
D$_2$O       & 1020 & 2551 & 0.000292 & 137 & 2.7 $\cdot 10^{-6}$ \\
O$_2$        & 1230 & 3076 & 0.000242 & 54  & 0.64 $\cdot 10^{-6}$ \\
CO$_2$       & 1560 & 3901 & 0.000191 & 85.5 & 6.0 $\cdot 10^{-6}$ \\
C (Diamond)  & 3520 & 8803 & 0.000085 & 290 & 45 $\cdot 10^{-6}$ \\
Be           & 1850 & 4627 & 0.000161 & 235 & 68 $\cdot 10^{-6}$ \\
\end{tabular}
\end{ruledtabular}
\end{center}
\end{table}

\subsection*{Validity of the model at high energy}

At high velocities neutrons can see individual nuclei, and the
form of the potential we assumed is not valid. This happens when
the wavelength of the neutron is smaller than the inter-atomic
distances. So we use the following limit for our model:
\begin{equation}
\lambda_{\text{min}} = \frac{2\pi}{k_{\text{max}}} = d
\end{equation}
where $d$ is the mean inter-atomic distances, and $k_{\text{max}}$ is
the maximum neutron wave vector allowed by our phenomenological
model. The table shows this limit for various materials.
\begin{center}
\begin{tabular}{ccccccc}
%\begin{ruledtabular}
\hline\hline
Nanoparticle & D$_2$ & D$_2$O & O$_2$ & CO$_2$ & C & Be \\
\hline
$d$ (nm)   & 0.32 & 0.32 & 0.35 & 0.36 & 0.18 & 0.20 \\
$k_{\text{max}}$ (nm$^{-1}$)   & 19 & 19 & 16 & 17 & 35 & 31 \\
\hline\hline
%\end{ruledtabular}
\end{tabular}
\end{center}

As we will see, this constraint is not of actual importance in the
range of parameters concerned, corresponding to the optimal
conditions of neutron thermalization.

There is another limit at high velocities. The neutron can excite the internal degrees of freedom of nanoparticles, so that the collision can be inelastic. We expect that the probability of phonon excitation is low because the phonon wavelenght should be in this case shorter than the considered size of nanoparticles. Also this process can only increase the efficiency of the neutron cooling.

\subsection*{Scattering amplitude within Born approximation}

Our goal is to calculate the three relevant quantities describing
the collision of a neutron with a nanoparticle: the total
scattering cross section, the absorption cross section, and the
mean energy loss. Let us note that in this problem the absorption
probability is extremely small with respect to the elastic
scattering one. Therefore the total interaction cross section is
approximately equal to the elastic one.

In the first Born approximation we can easily compute this
scattering amplitude in the center-of-mass system (c.m.s.), that
is, the amplitude for a neutron with incident wave vector
$\vect{k}$ to be scattered at final wave vector $\vect{k}'$:
\begin{equation}
f(\theta) = - \frac{1}{4 \pi} \frac{2m}{\hbar^2} \int e^{i
(\vect{k} - \vect{k}')\cdot \vect{r}} V(\vect{r}) d\vect{r}
\end{equation}
$\theta$ being the angle between $\vect{k}$ and $\vect{k}'$. Let
$\vect{q} = \vect{k} - \vect{k}'$ be the momentum transfer. The
collision is elastic in the c.m.s. so that $k = k'$ and
\begin{equation}
q = 2 k \sin(\theta / 2).
\end{equation}
Finally, we find
\begin{equation}
\label{amplitude}
f(\theta) = - \frac{2m}{\hbar^2} V \ R^3 \frac{1}{(q R)} j_1(q R)
\end{equation}
where $j_1(X)$ is the first spherical Bessel function:
\begin{equation}
j_1(X) = \frac{\sin(X)}{X^2}-\frac{\cos(X)}{X}.
\end{equation}

%\subsection*{Elastic cross section}

>From the scattering amplitude we can calculate the elastic cross
section:
\begin{equation}\label{xsel}
\sigma_{s} = \int \left| f \right|^2 d\Omega =  2 \pi
\left|\frac{2m}{\hbar^2} V \right|^2 \ R^6 \frac{1}{(k R)^2} I(k
R),
\end{equation}
where
\begin{eqnarray}\label{Auxint}
I(k R) & = & \int_0^{2kR} \frac{1}{x}j_1(x)^2 dx \nonumber \\
                 & = & \frac{1}{4}\left(1-\frac{1}{(2 k R)^2} +
                 \frac{\sin(4 k R)}{(2 k R)^3} - \frac{\sin^2(2 k R)}{(2 k R)^4}\right).
\end{eqnarray}

%\subsection*{Absorption cross section}

By using the optical theorem, we can calculate the absorption
cross section $\sigma_{a}$ at the first order Born approximation:
\begin{equation}
\text{Im}(f(\theta = 0)) = \frac{k}{4\pi} \sigma_{a}
\end{equation}
As $f(0) = -\frac{1}{3} \frac{2m}{\hbar^2} V R^3$, one obtains:
\begin{equation}
\label{xsabs}
\sigma_{a} = \frac{4 \pi}{3} \ \frac{2m}{\hbar^2} V_1 \ R^4 \ \frac{1}{k R}.
\end{equation}

\begin{figure}
\includegraphics[angle=270,width=0.9\linewidth]{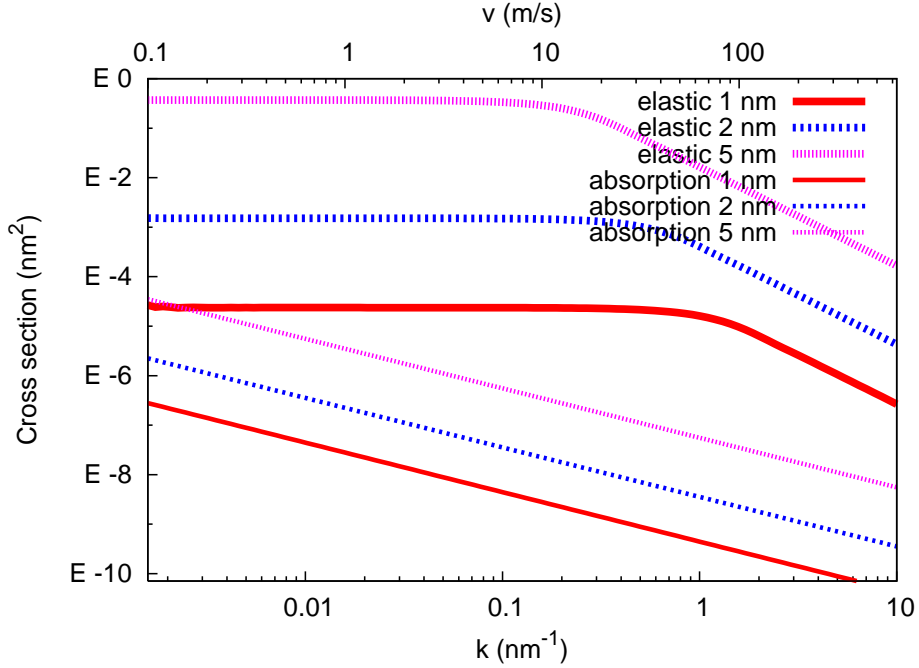}
\caption{Elastic and absorption cross sections as a function of
neutron velocity, for three values of the deuterium nanoparticles
radii: 1, 2, and 5 nm. } \label{xsections}
\end{figure}

 Calculated elastic and absorption cross sections are presented
in Fig.~\ref{xsections} as a function of neutron velocity. The
calculations are performed for different nanoparticle's radii. For
low energies where $kR \ll 1$, one can easily see from this figure
as well as from (\ref{xsel}) and (\ref{xsabs}) that $\sigma_{s}
\sim R^6$ and $\sigma_{a} \sim R^3$.

\subsection*{Mean energy loss per collision}

We consider a collision between the neutron and a nanoparticle
which has a mass $A$, in units of the neutron mass. Let $\theta$
be the scattering angle. As the mass of the nanoparticle is much
greater than the mass of the neutron, we know that the energy
transfer, in the first order in $1/A$ is given by:
\begin{equation}
\label{DeltaE}
\frac{\Delta E}{E} \simeq - 4 \frac{1}{A} \sin(\theta/2)^2
\end{equation}
Since we know the scattering amplitude at the Born approximation,
we can calculate the mean relative energy loss:
\begin{equation}
\frac{\left\langle \Delta E \right\rangle}{E} = \frac{\int
\frac{\Delta E}{E} \frac{d\sigma}{d\Omega} d\Omega}{\int
\frac{d\sigma}{d\Omega} d\Omega}
 = \frac{2\pi}{\sigma_{s}} \int_0^\pi \frac{\Delta E}{E}(\theta)
 \left|f(\theta)\right|^2 \sin{\theta} d\theta
\end{equation}
In terms of the variable $x = q R = 2 k R \sin(\theta/2)$:
\begin{equation}
\frac{\Delta E}{E} = -\frac{1}{A} \frac{x^2}{(kR)^2} \ \text{ and
} \ f = - \frac{2m}{\hbar^2} V R^3 \frac{1}{x} j_1(x)
\end{equation}
Now we can express the mean relative energy loss:
\begin{equation}
\label{DeltaEmoy}
\xi = \frac{\left\langle \Delta E \right\rangle}{E} =
- \frac{1}{A} \frac{1}{(kR)^2} \frac{J(kR)}{I(kR)}
\end{equation}
with
\begin{equation}
\frac{J(kR)}{I(kR)} = \frac{\int_0^{2k R} x j_1(x)^2
dx}{\int_0^{2k R} \frac{1}{x} j_1(x)^2 dx}.
\end{equation}
The first integral in this expression can be expressed in terms of
special functions but has to be calculated numerically where the
second one was calculated previously in (\ref{Auxint}).

\begin{figure}
\includegraphics[angle=270,width=0.9\linewidth]{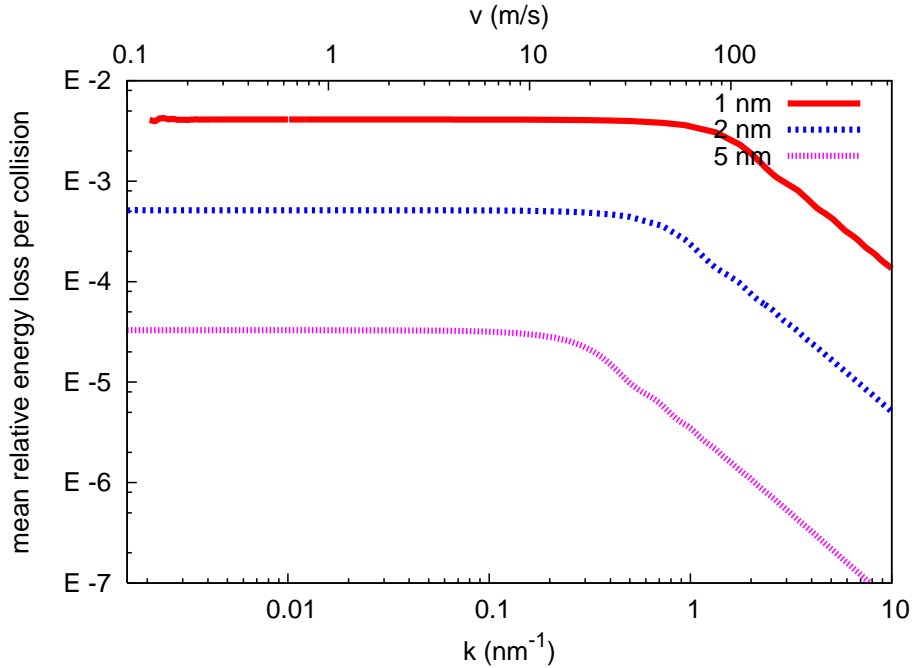}
\caption{Relative mean energy loss per collision $\xi$, as a
function of neutron velocity, for three values of the deuterium
nanoparticles radii: 1, 2, and 5 nm. } \label{xi}
\end{figure}

The relative mean energy loss as a function of neutron velocity is
presented in Fig.~\ref{xi} for different values of nanoparticles radii. Let us note that, for small energies where $kR \ll 1$, this
loss behaves as $\xi \sim R^{-3}$.

\subsection*{Interpretation of the results}

As expected, the absorption cross section follow a $1/v$ law, and
is proportional to the nanoparticle's mass $A_0 R^3$. At low
velocities, the elastic cross section and the energy loss are
constant up to some critical momentum $k_c \approx 1/R$ beyond
which the collision becomes anisotropic. The total elastic
cross section and the energy loss therefore decrease rapidly.

As we have mentioned previously, the probability of collision is
strongly dependent on the size of the nanoparticle, as $R^6$: due
to the coherent scattering of a neutron at nuclei in the
nanoparticle, the scattering cross section is proportional of the
square of the number of nuclei in the nanoparticle. From the
condition $k_c \approx 1/R$, the critical momentum is higher for
smaller nanoparticles.

The efficiency of the moderation results from competition
between the absorption process and the elastic scattering. The
probability of the first process is proportional to the absorption
cross section $\sigma_{a}$ multiplied by the number of collisions
needed to slow down neutrons (this number is inversely
proportional to the averaged energy losses $\xi$). The probability
of elastic scattering is proportional to the elastic cross section
$\sigma_{s}$. As we have emphasized, all these three values
$\sigma_{a}$, $\xi$, $\sigma_{s}$ are strongly $R$ dependent but
their combination $\sigma_{a}/\xi\sigma_{s}$ which defines the
moderator efficiency is not. This combination depends only on the
product $kR$ and has its minimum for $kR \sim 1$ (as it can be
seen from Fig. \ref{xsections} and Fig. \ref{xi}). Therefore it is
natural to expect the best moderator properties for neutron
velocities close to the critical one $k_c \sim 1/R$. This ratio
decreases for both higher and lower velocities. We will see that a
more detailed analysis in section \ref{infmoderator} will confirm
this general statement.

\subsection*{Validity of the first Born approximation}

Sufficient condition for the validity of the first Born
approximation is:
\begin{equation}
\frac{2 m}{\hbar^2} V_0 R^2 \ll 1
\end{equation}
The lower the radius, the better the approximation. To see where
this condition is verified, let us calculate the maximum radius
$R_{\text{Born}}$ to have an accuracy better than $10 \%$:
$\frac{2 m}{\hbar^2} V_0 R_{\text{Born}}^2 = 0.1$.

\begin{center}
\begin{tabular}{ccccccc}
%\begin{ruledtabular}
\hline\hline
Nanoparticle & D$_2$ & D$_2$O & O$_2$ & CO$_2$ & C & Be \\
\hline
$R_{\text{Born}}$ (nm)   & 4.5 & 3.7 & 5.4 & 4.5 & 2.6 & 2.9 \\
\hline\hline
%\end{ruledtabular}
\end{tabular}
\end{center}

Actually, the approximation works much better, because our
potential is repulsive \cite{Taylor}. A more accurate calculation,
with partial waves given in Appendix \ref{partwave} shows that the
accuracy of the Born approximation is sufficient enough for our
needs.

%-----------------------------------------------------------%
\section{\label{infmoderator}Model of the infinite moderator}
%-----------------------------------------------------------%

The role of any moderator is to increase the density of neutrons
in the phase space. In this section, we will determine the
influence of the properties of nanoparticles -- their material size -- on the evolution of neutron density. In order to
achieve that, we have to identify the criteria for cooling efficiency.

The most general problem that we perceive is the following.
We have a moderator medium made of free nanoparticles in
suspension in superfluid helium. There is a neutron source, at a
certain location, with a certain velocity distribution. What is the neutron density at every point in the moderator medium and what is the distribution of velocities? The quantity describing the
state of the system in this problem is a seven-variable function,
the density of the neutrons in the phase space $n(\vect{r},
\vect{k}, t)$. In general case, this density evolves according to the transport equation, or Boltzmann equation, and all
the variables can be coupled.

Solution of this equation in general case can be quite complicated
task so we will propose a first necessary criterium of cooling
efficiency within the model of infinite moderator. In this model,
we eliminate the space variable in the density; we assume that the
sources are uniformly distributed, and that the medium is
homogeneous and isotropic. The quantity which describes the system
is only a function of the energy (and also of time, but it is only the
stationary regime that interests us). We can analytically compute the
energy spectrum of the neutrons in this moderator, using the three
quantities $\sigma_s$, $\sigma_a$, $\xi = \frac{\left\langle
\Delta E\right\rangle}{E}$ describing the elementary processes.

\subsection*{The energy spectrum in the infinite moderator}

To compute the neutron energy spectrum in the moderator $n(E, t)$,
we have to use the equation for the conservation of the number of
neutrons:
\begin{equation}
\drond{n}{t} - \drond{q}{E} + \text{Absorption} - \text{Source} = 0
\end{equation}
where $q(E, t)$ is the cooling current, i.e. the number of
neutrons scattered from an energy greater than $E$ to an energy
lower than $E$, in the unit of time. The absorption term can be
related to the macroscopic absorption cross section $\Sigma_a = N
\sigma_a$ via the flux variable $\phi(E, t) = v n(E, t) =
\sqrt{\frac{2 E}{m}} n(E, t)$, so that the conservation equation,
using the flux variable, can be expressed as:
\begin{equation}
\label{conservation}
\sqrt{\frac{m}{2 E}} \drond{\phi}{t} - \drond{q}{E} + \Sigma_a \phi = \text{Source}
\end{equation}
Now we have to specify the cooling current under the assumption
that $\xi \ll 1$ which is clear to be a very good approximation
for neutron scattering on nanoparticles (see Fig.~\ref{xi}).
Under this assumption the cooling current has the simple and natural
form \cite{react}:
\begin{equation}
\label{q} q(E) = E \xi(E) \Sigma_s(E) \phi(E).
\end{equation}

%\subsection*{Solution in stationary regime}

Let us consider the stationary regime. The conservation equation is
completely expressed in terms of our three microscopic quantities,
and so the time independent equation is given by
\begin{equation}
-\deriv{}{E} \left[\xi \Sigma_s E \phi \right] + \Sigma_a \phi = 0
\end{equation}
We have not expressed the source term, because if the source is
punctual (a given flux $\Phi_0$ at a given energy $E_0$) then we
can solve the equation without any source in the domain $0 < E <
E_0$ and the source is a boundary condition: $\phi(E_0) = \phi_0$.
So now, as we have a usual first order linear differential
equation, we can solve it analytically with the previous boundary
condition:
\begin{equation}
\phi(E) = \phi_0 \frac{E_0 \xi (E_0)\Sigma_s(E_0)}{E \xi(E)
\Sigma_s(E)} \ \text{exp} \left( -\int_E^{E_0} \frac{\Sigma_a}{\xi
\Sigma_s} \frac{d\epsilon}{\epsilon} \right)
\end{equation}
We can notice that, if there is no absorption, and if both $\xi$
and $\Sigma_s$ are energy independent, we find the usual behavior, the so called the $1/E$ flux law.

From the flux, it is useful to derive the density $n(k)$ of
neutrons in the phase space:
\begin{equation}
n(k) = n(k_0) \left( \frac{k_0}{k}\right)^2 \frac{( \xi
\Sigma_s)(k_0)}{(\xi \Sigma_s)(k)} \ \text{exp}\left(- 2
\int_k^{k_0} \frac{\Sigma_a}{\xi
\Sigma_s}\frac{d\kappa}{\kappa}\right)
\end{equation}

\subsection*{A necessary condition for efficient cooling}

\begin{figure}
\includegraphics[angle=270,width=0.9\linewidth]{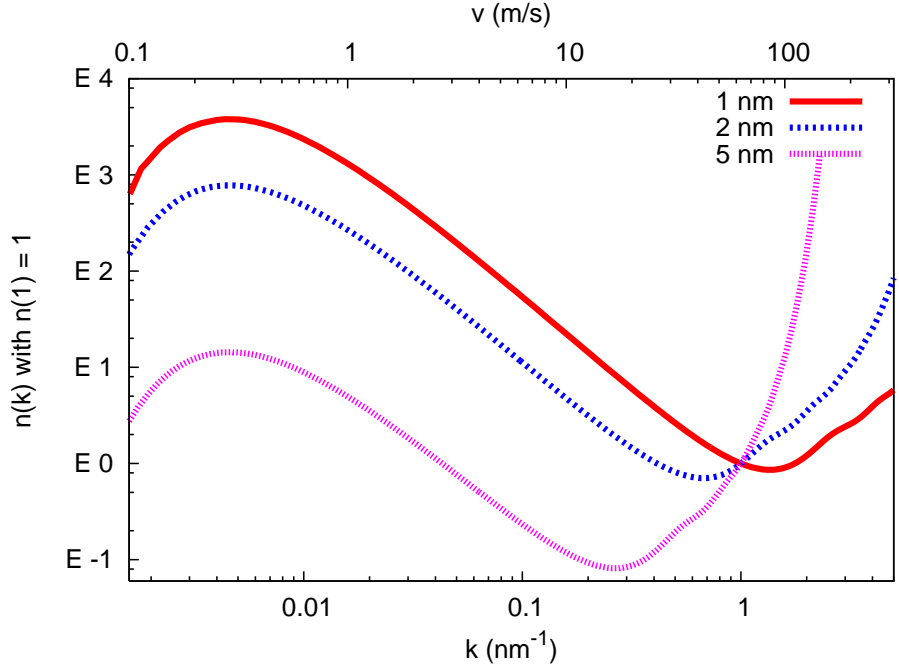}
\caption{Phase space density in stationary regime as a function of
velocity, for three radii of deuterium nanoparticles. The source
is assumed to be monochromatic, with initial energy
equal/greater than the maximum energy presented in the graph.}
\label{density}
\end{figure}

The density of neutrons in the velocity space, for the stationary
regime in the infinite moderator, is presented in Fig. \ref{density}.

To interpret this figure, we have to assume that there is a
monochromatic neutron source at velocity $k_0$, and we extract the
neutrons at the velocity $k_1 < k_0$. Then the relative increase
of density in the phase space is given by $n(k_1)/n(k_0)$. As the
goal is to increase the density of neutrons in the phase space, we
see that the curve gives us a necessary condition for the
efficiency of the cooling. Indeed, the cooling is efficient only
if $n(k_1) > n(k_0)$, that is in the decreasing part of the curve.
This differential necessary condition is then given by
$\deriv{\ln{n}}{k} (k_0)<0$, that we can put into the form:
\begin{equation}
\frac{\Sigma_a}{\xi \Sigma_s} < 1 + \frac{1}{2} \deriv{\ln(\xi \Sigma_s)}{\ln{k}}
\label{condition}
\end{equation}

\begin{figure}
\includegraphics[angle=270,width=0.9\linewidth]{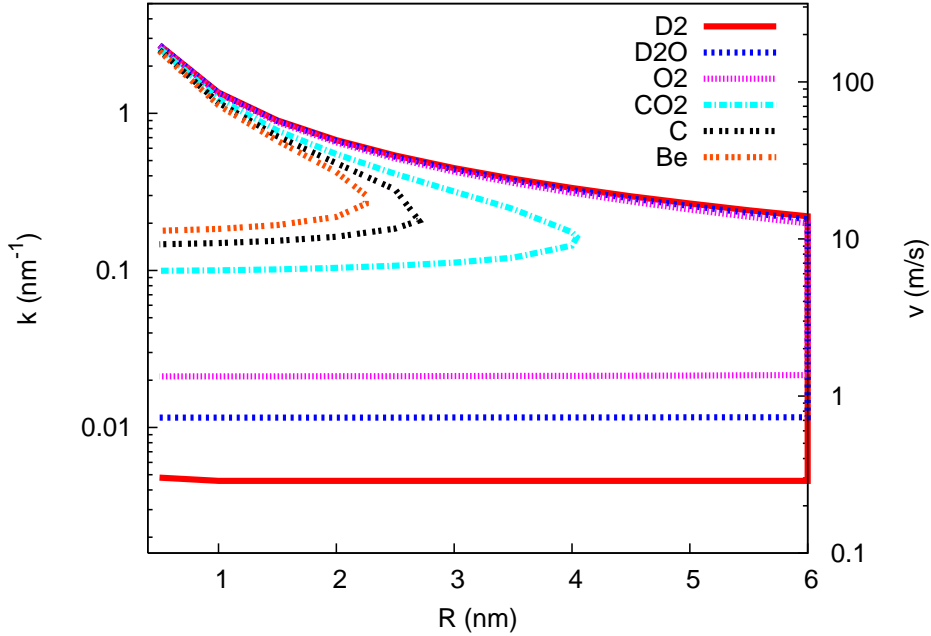}
\caption{Efficiency domain of the moderation as a function of
nanoparticle's radii, for different materials. Interference
effects are not taken into account.} \label{coolingDomain}
\end{figure}

The cooling domain is the velocity domain in which this condition
is satisfied. Figure \ref{coolingDomain} shows that the cooling is
not efficient both at low velocities and at high velocities.

At low velocities, the cooling is not efficient because the
absorption is important compared to the diffusion cross section
and the energy loss, and at high velocities (basically at
velocities higher than the critical velocity), the cooling is not
efficient because both the scattering cross section and the energy
loss fall down. Figure \ref{coolingDomain} shows the cooling
domain as a function of the radius of the nanoparticles, for
several materials. These curves leads to three remarks. For
certain materials, such as diamond, and for certain radius of
nanoparticles, there is no cooling domain at all. The cooling
domain is bigger for smaller nanoparticles, because for smaller
nanoparticles, the critical velocity is higher. The domain for
deuterium nanoparticles include the domain for all the other
materials.

\begin{figure}
\includegraphics[angle=270,width=0.9\linewidth]{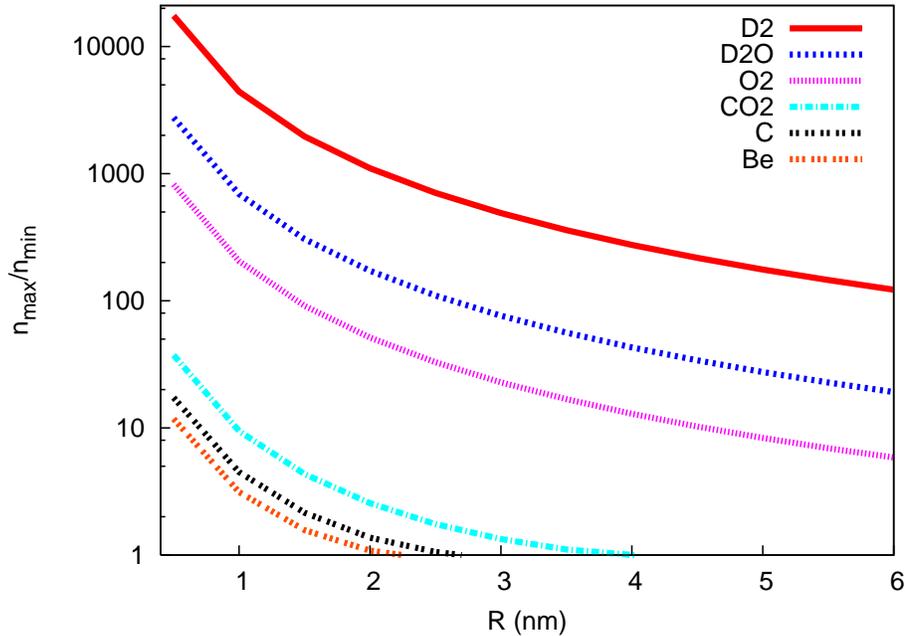}
\caption{\label{maxdensity}Maximum compression of phase space
density possible, as a function of the radius of the nanoparticles, and
for different materials. Interference effects are not taken into
account.}
\end{figure}

We can now put a limit for the total increase of density in the
phase space achievable for the infinite moderator and for a
monochromatic source. Indeed, in Fig. \ref{density} we see that
the best we can do is to have a source with $k_0$ at the minimum
of the curve, and to extract neutrons with $k_1$ at the maximum of
the curve. The maximum gain is given by $n(k_1)/n(k_0)$. Figure
\ref{maxdensity} shows this limit, as a function of the radius of
the nanoparticles, and for different materials. We see that this
maximum is bigger for smaller nanoparticles, because the cooling
domain is higher. The best material for this criterium is
deuterium.

%------------------------------------------------------%
\section{\label{realmoderator}More realistic moderators}
%------------------------------------------------------%

The consideration of the infinite moderator gives us a necessary
condition for the efficiency of the cooling. The results presented
do not depend on the density of nanoparticles in the moderator.
When it comes to realistic moderators, we have to take two
characteristics of the moderation into account: the size of the moderator, and
the thermalization time. These two quantities depend on the
density of nanoparticles, and for the calculations we assumed that
in a given volume, the total mass of nanoparticles represents 1 \%
of the total mass of helium. That means that for a gel of
smaller nanoparticles, the distance between the nanoparticles is
assumed to be smaller.

Let us pursue the idea to estimating the size of the moderator and
the thermalization time. Suppose a monochromatic neutron source at velocity $k_0$ inside the efficient domain, and
that we want to increase the density in phase space by the factor
$e$. According to Fig. \ref{density}, we have to extract neutrons
in the moderator with an energy $k_0-\Delta k$, where
$\Delta k$ corresponds to an increase of factor $e$ for the
function $n(k)$:
\begin{equation}
\Delta k = \left[\deriv{\ln{n(k)}}{k} (k_0)\right]^{-1}.
\label{deltak}
\end{equation}
We know that this decrease of velocity by $\Delta k$ corresponds
to a certain number of collisions $\Delta N$
\begin{equation}
\Delta N = \frac{1}{\xi} \frac{\Delta k}{k_0}
\label{deltan}
\end{equation}
During $\Delta N$ collisions, neutron assume a Brownian-like trajectory, and we define $L(k_0)$ as the square-mean-root of the distance travelled. We also define the thermalization time
$\tau(k_0)$ as the time it takes for the neutron to collide
$\Delta N$ times.

\subsection*{Estimation of the moderator's size}

The square mean distance travelled by a neutron $\Delta r^2$ after
$\Delta N$ collisions is given by \cite{Barjon}:
\begin{equation}
\Delta r^2 = \frac{4}{A} \frac{1}{\xi \Sigma_s^2} \Delta N
\end{equation}
This expression was obtained under the assumption that the
diffusion cross section $\Sigma_s$ and the mean relative energy
loss $\xi$ are energy-independent, and that the scattering is
nearly isotropic.

From equations (\ref{deltak}) and (\ref{deltan}), we can estimate the
characteristic length $L(k)$ of the moderator, corresponding to an
increase of factor $e$ of the phase space density:
\begin{eqnarray}
\frac{1}{L(k)^2} & = & \deriv{\ln(n)}{r^2} = \frac{A}{4}
(\xi \Sigma_s)^2 \deriv{\ln(n)}{\ln(k)} \nonumber \\
& = & \frac{A}{2} (\xi \Sigma_s)^2 \left[1 + \frac{1}{2}
\deriv{\ln(\xi \Sigma_s)}{\ln(k)} - \frac{\Sigma_a}{\xi
\Sigma_s}\right]
\end{eqnarray}
Note that we find our necessary condition (\ref{condition}), in the
sense that the condition is satisfied if and only if the size $L$
is real.

\begin{figure}
(a)
\includegraphics[angle=270,width=0.85\linewidth]{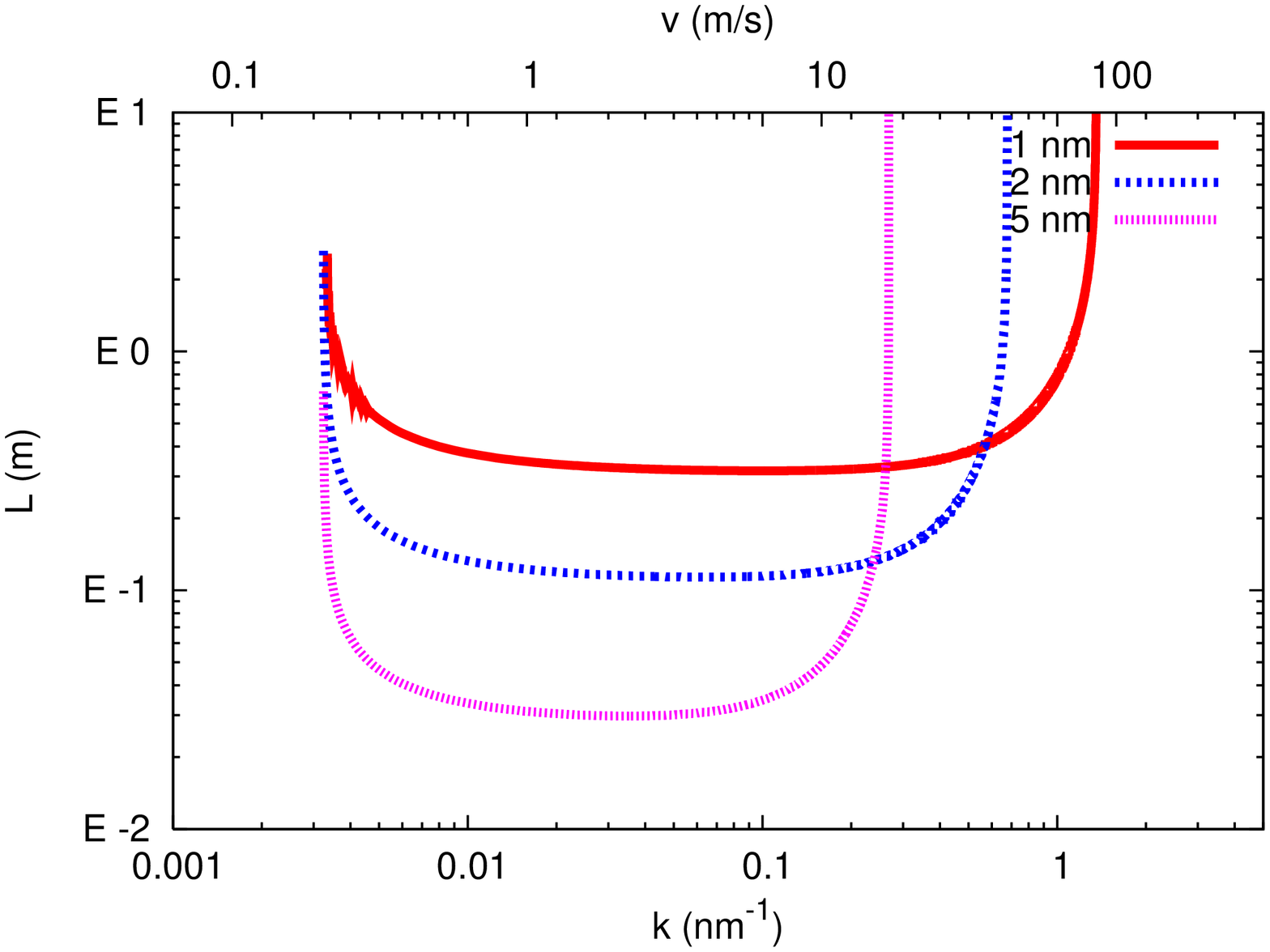}

\noindent (b)
\includegraphics[angle=270,width=0.85\linewidth]{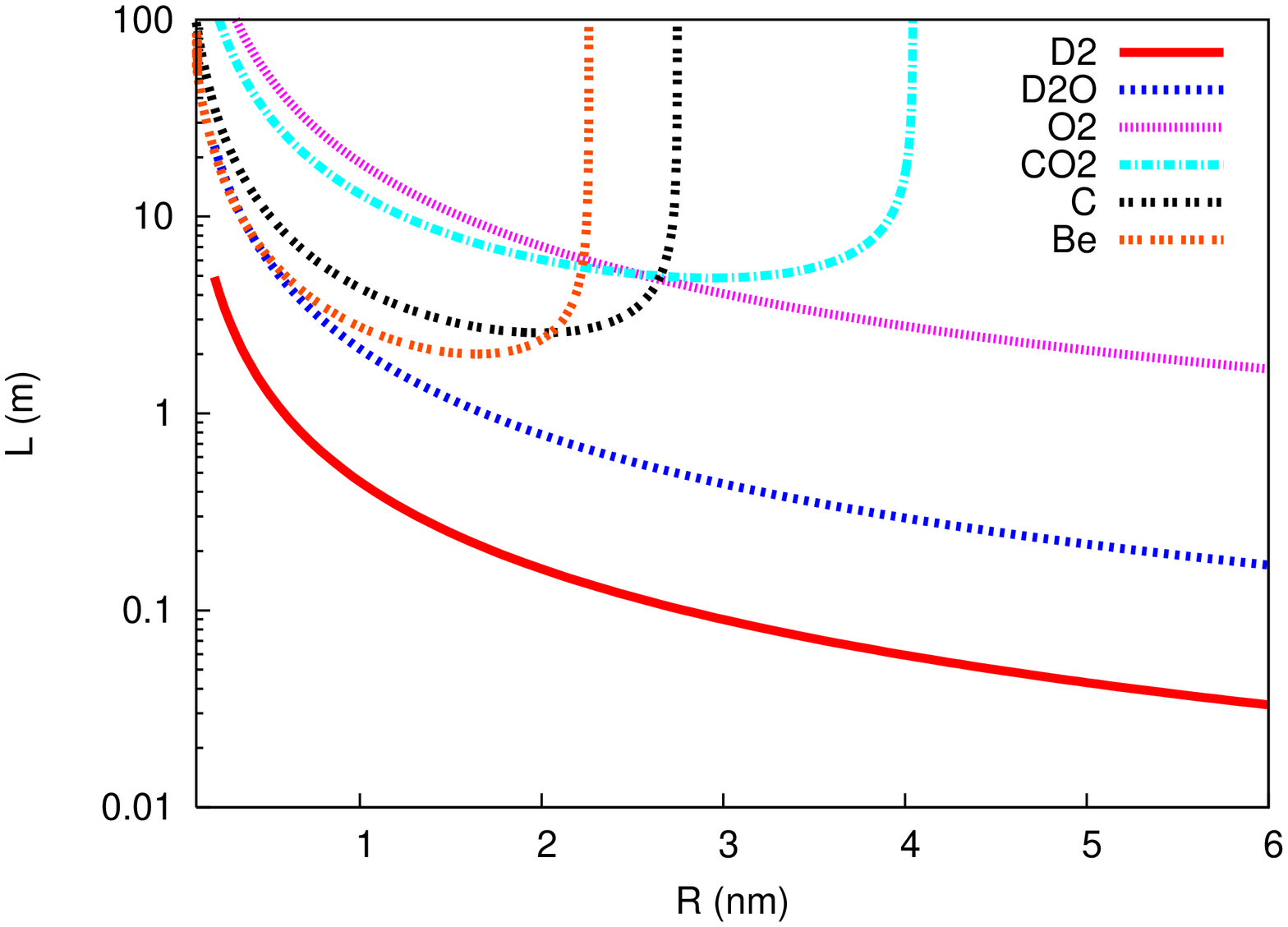}
\caption{Moderator size needed to increase the density in phase
space by a factor $e$. Interference effects are not taken into
account. (a) Moderator size as a function of velocity, for three
radii of deuterium nanoparticles. (b) The minimum moderator size
as a function of the nanoparticles radius, for different
materials.} \label{size}
\end{figure}

\subsection*{Estimation of the thermalization time}

Now we must estimate of the thermalization time. We know that
the mean time between two collisions is given by:
\begin{equation}
\tau_{\text{coll}} = \frac{m}{\hbar} \frac{1}{k \Sigma_s}.
\end{equation}
The time needed for a neutron to collide $\Delta N$ times
is therefore simply $\tau_{\text{coll}} \Delta N$. From equations (\ref{deltak}) and
(\ref{deltan}), we can estimate the thermalization time $\tau(k)$,
corresponding to an increase by a factor $e$ of the phase space
density:
\begin{eqnarray}
\frac{1}{\tau(k)} & = & \frac{\hbar}{2 m} k^2 \xi \Sigma_s \deriv{\ln(n)}{k} \nonumber \\
& = & \frac{\hbar k}{2 m} \xi \Sigma_s \left[1 + \frac{1}{2}
\deriv{\ln(\xi \Sigma_s)}{\ln(k)} - \frac{\Sigma_a}{\xi
\Sigma_s}\right]
\end{eqnarray}
Again, we find our necessary condition (\ref{condition}), in the
sense that the condition is satisfied if and only if the
characteristic time $\tau$ is positive.

\begin{figure}

(a)
\includegraphics[angle=270,width=0.85\linewidth]{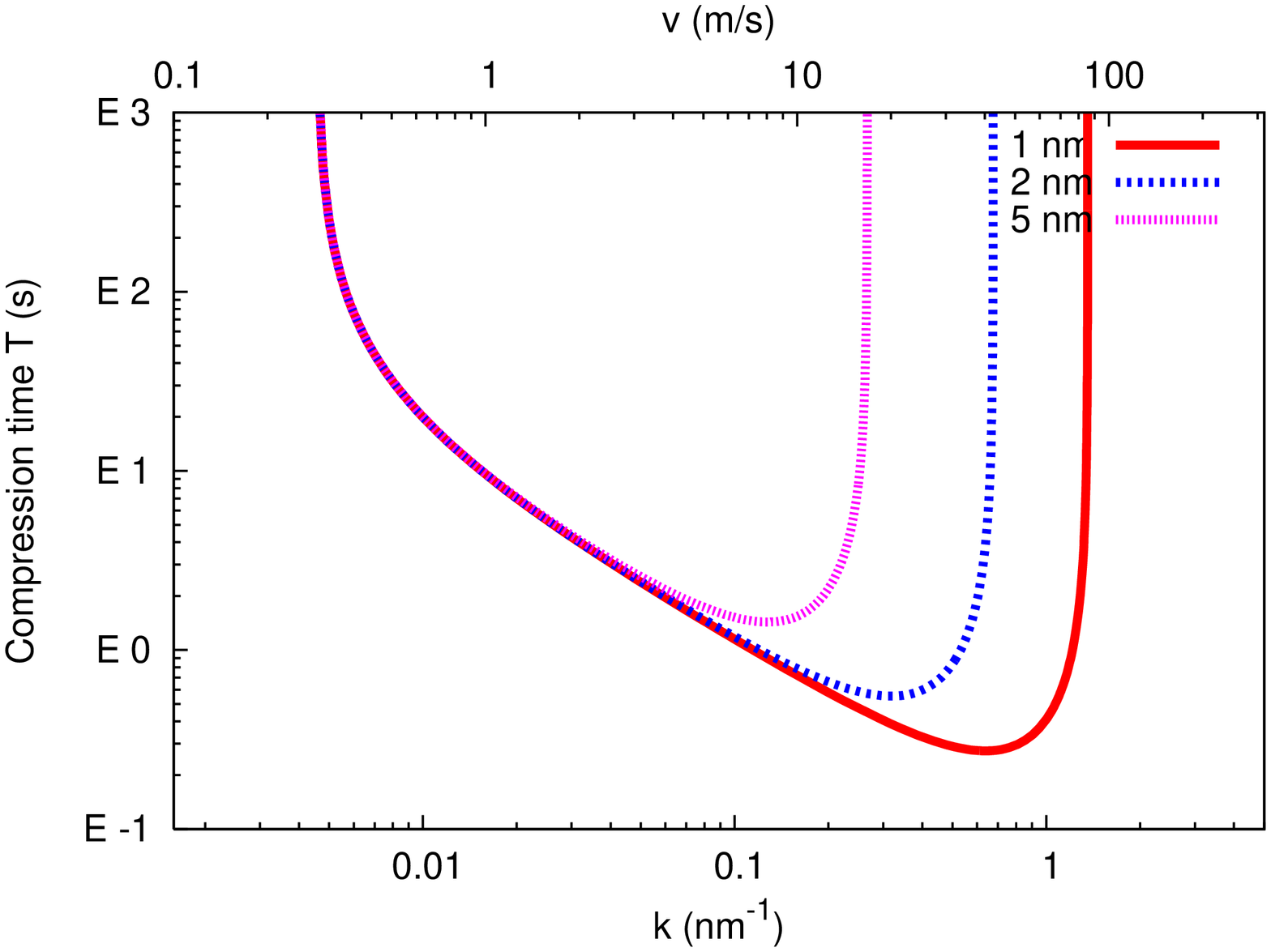}

\noindent (b)
\includegraphics[angle=270,width=0.85\linewidth]{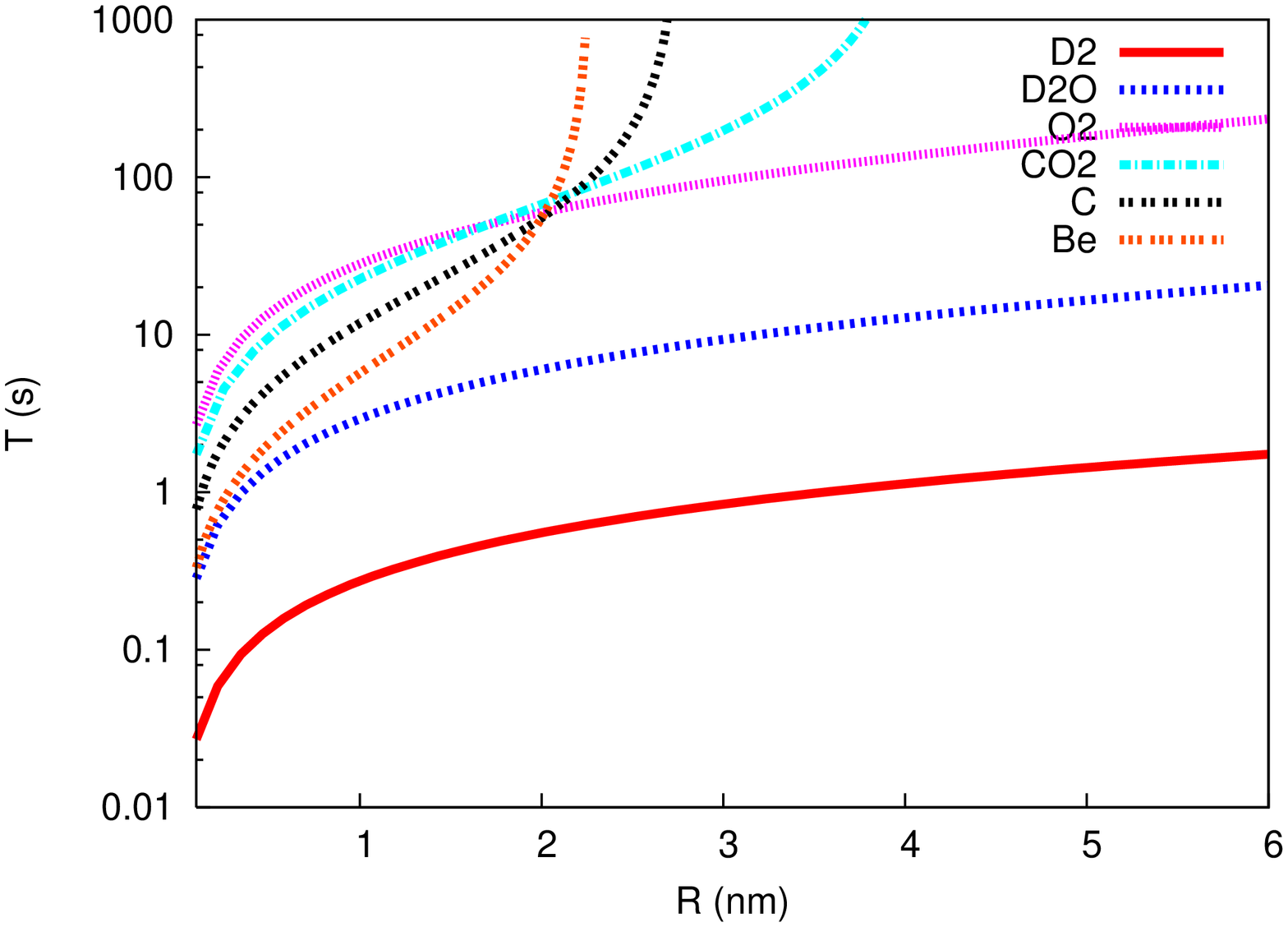}
\caption{Thermalization time needed to increase the density in
phase space by a factor $e$. Interference effects are not taken into
account. (a) Thermalization time as a function of velocity, for
three radii of deuterium nanoparticles. (b) The minimum
thermalization time as a function of the nanoparticles radius, for
different materials.} \label{time}
\end{figure}

\subsection*{Comments on the results}

The results are shown in Fig. \ref{size} and Fig. \ref{time}. Let
us first comment on the figures for the deuterium example. We see first that outside of the efficiency domain of the cooling, both the thermalization time and the moderator size diverge. We also
remark that the moderator size is nearly constant over the
efficiency domain, although the thermalization time presents a
clear minimum near the critical velocity. But the main difference
between the two conditions is the dependence on the radius of the
nanoparticles. For small nanoparticles, we need fewer collisions to
cool down the neutrons, and we can see in Fig. \ref{time} that the
smaller the nanoparticles, the smaller the thermalization time.
On the other hand, for small nanoparticles, the mean free
path is high, and we see in Fig. \ref{size} that the smaller the
nanoparticles, the bigger the moderator size. We cannot therefore optimize the radius of the nanoparticle in a general way; we must look at more practical considerations.

We can use both the size of the moderator and the thermalization
time as criteria to compare materials. We can see that using both
criteria, deuterium is better than any other material by almost
one order of magnitude. We also see that the absolute values are
very competitive. Indeed, to increase the density in phase space
by factor $e$, the characteristic size is a few tens of
centimeters and the thermalization time is less than one second,
for a typical radius of 2 nm.

\subsection*{Practical proposal}

\begin{figure}
\includegraphics[width=0.9\linewidth]{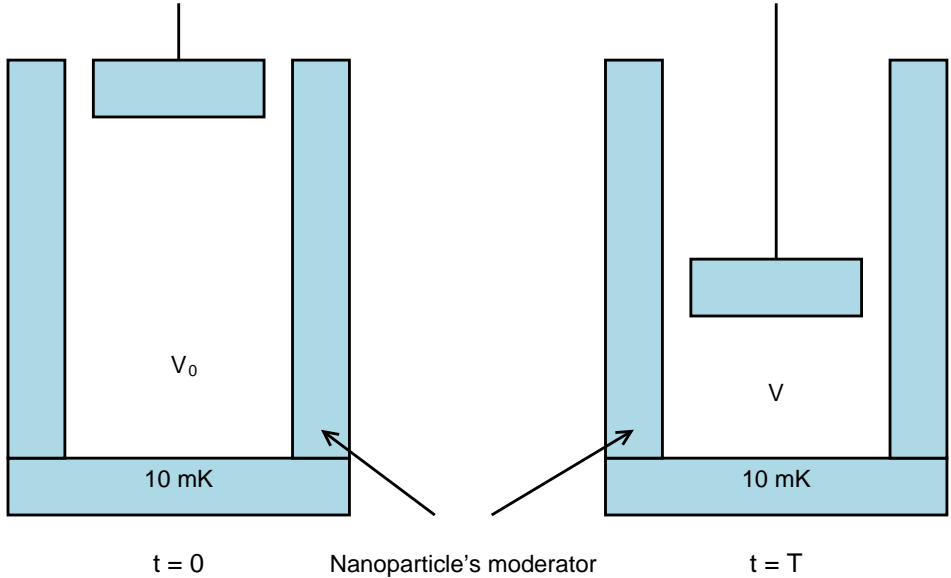}
\caption{Practical proposal to increase the neutron density in
phase space, combining a compression in the real space with a
piston as well as a compression in the velocity space with the
nanoparticle's moderator. } \label{practical}
\end{figure}

\begin{figure}
\includegraphics[angle=270,width=0.9\linewidth]{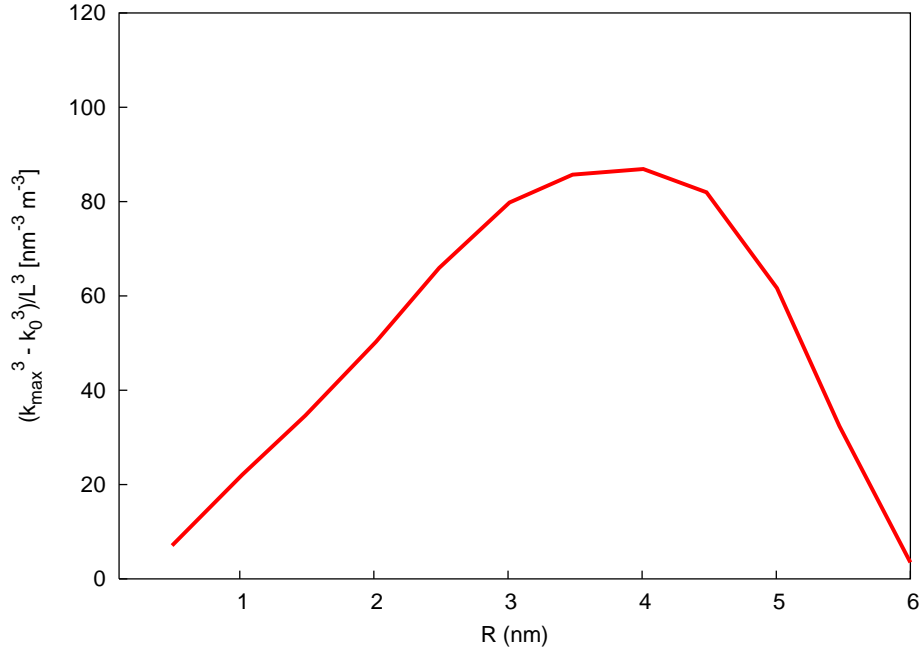}
\caption{Practicle efficiency criterium : effective cooling phase space volume $k_{\text{max}}^3 - k_0^3$ over characteristic volume of the moderator $L^3$. This criterium is plotted for deuterium nanoparticles as a function of the radius of the nanoparticle.} \label{criterium}
\end{figure}

To achieve a compression in the phase space by many orders of
magnitude, we can associate the nanoparticle's moderator to a
compression in real space. To explain this idea in more
details, Fig. \ref{practical} may be of help. We start with a gas of
neutrons uniformly located in the initial volume $V_0$ with a
velocity distribution with a mean value $v_0$ and a width $\Delta
v$. We decrease the volume by a factor $V/V_0$. According to the
Liouville theorem, the width of the velocity distribution will
increase as $\Delta v \longrightarrow (\frac{V_0}{V})^{1/3} \Delta
v$, so that the density in the velocity space decreases by the same
factor (in absence of absorption this relation would be precise, it is approximately valid in the domain of efficient cooling of neutrons defined above). This is the case when there is no moderator, but if we
put a nanoparticle's moderator in the volume (not in all the
volume, but in a little box in a corner), the moderation will
compensate the increase of density in the velocity space. So finally the maximum density in the phase space achievable is the product of the initial density, the volume compression
$V_0/V$, and the maximal compression due to the moderator shown in
Fig. \ref{maxdensity}. As we can in principle reach as much volume
compression as we wish, there is no theoretical limitation
of the maximum compression in phase space. With this method, it is
not necessary to use all the cooling domain shown in Fig.
\ref{coolingDomain} (two orders of
magnitude of velocity in a favorable case); we can concentrate the process around the
optimal velocity. This optimal velocity should be chosen according to two practical considerations:
\begin{enumerate}
    \item We are limited by the size of the moderator, which should not exceed a few tens of centimeters for practical purposes.
    \item The optimal velocity should correspond to a temperature
          greater than that achievable with a dilution cryostat, which
          is about $10$ mK. The corresponding wave vector is about $k_0 = 0.2$ nm$^{-1}$.
\end{enumerate}
This two requirements are in competition, so we propose a practical criterium defined as :
\begin{equation}
\frac{\Delta k^3}{L^3} = \frac{k_{\text{max}}^3 - k_0^3}{L^3}.
\end{equation}
It is the ratio between the phase space volume for which we can have an efficient cooling in an actual device and the characteristic volume of the moderator. This criterium is plotted in Fig. \ref{criterium} for deuterium nanoparticles as a function of the size of the nanoparticle. This figure shows that the interresting range for nanoparticles radius is between two and five nanometers.
Althought the thermalization time is of principle importance, we showed that for deuterium nanoparticle this time is in any relevant cases small compared to the neutron's lifetime.

%-------------------------------%
\section{Extensions of the model}
%-------------------------------%

For this special case of moderator nanoparticle gels there are several limitations to the validity of our model. One obvious limitation is that chains of nanoparticles are not taken into account; in this paper we neglect such chains and consider only the cooling of neutrons caused by the collisions with free nanoparticles. We have already mentioned the limitation at high energies in section \ref{model}, this limitation is inherent to our assumptions about
the interaction of a neutron and a single nanoparticle. But there
is a limitation also at low velocities, which is specific to the
gel.

\subsection*{Limitation at low energy}

At low velocities, we cannot neglect the interferences between the
waves diffracted by several nanoparticles. This happens when the
wavelength of the neutron is bigger than the distance between the
nanoparticles. We see that this limitation does not come from our
assumptions about the interaction of a neutron with a single
nanoparticle. This limitation is specific to the practical medium
that we plan to use, that is, the gel of nanoparticles. Let $D$ be
the distance between the nanoparticles in the gel; we can compute
$D$, assuming that the total mass of nanoparticles is $1\%$ of
the total mass of helium: $D = R / N_0^{1/3}$. In the following
table we give $D$ for nanoparticles with radius $1$ nm. We also
give the minimal wave vector for the validity of the model
$k_{\text{min}}$, defined as follow: $\lambda_{\text{max}} =
\frac{2\pi}{k_{\text{min}}} = D$.
\begin{center}
\begin{tabular}{ccccccc}
%\begin{ruledtabular}
\hline\hline
Nanoparticle & D$_2$ & D$_2$O & O$_2$ & CO$_2$ & C & Be \\
\hline
$D_{\text{1 nm}}$ (nm)       & 8.7 & 15.1 & 16.0 & 17.4 & 22.7 & 18.4 \\
$k_{\text{min}}$ (nm$^{-1}$) & 0.7 & 0.4  & 0.4  & 0.4  & 0.3  & 0.4  \\
\hline\hline
%\end{ruledtabular}
\end{tabular}
\end{center}

We can see that this is a serious limitation, but not that
dramatic. Actually, for practical purposes, the lowest energy we
are interested in is that corresponding to the lowest temperature
achievable with a dilution cryostat, which is about $10$ mK. The
corresponding wave vector is about $0.2$ nm$^{-1}$. So we only
need to know , for practical purposes, the first correction of
$\sigma_s$, $\sigma_a$ and $\xi$ due to the interference of
neutron waves on the neighboring nanoparticles. And this can be
done, because the first correction can be estimated considering
only two nanoparticles. This approach is valid until the neutron
wavelength covers three or more nanoparticles, that means that
it is valid until $k_{\text{min}}$/2.

\begin{figure}
\includegraphics[angle=270,width=0.9\linewidth]{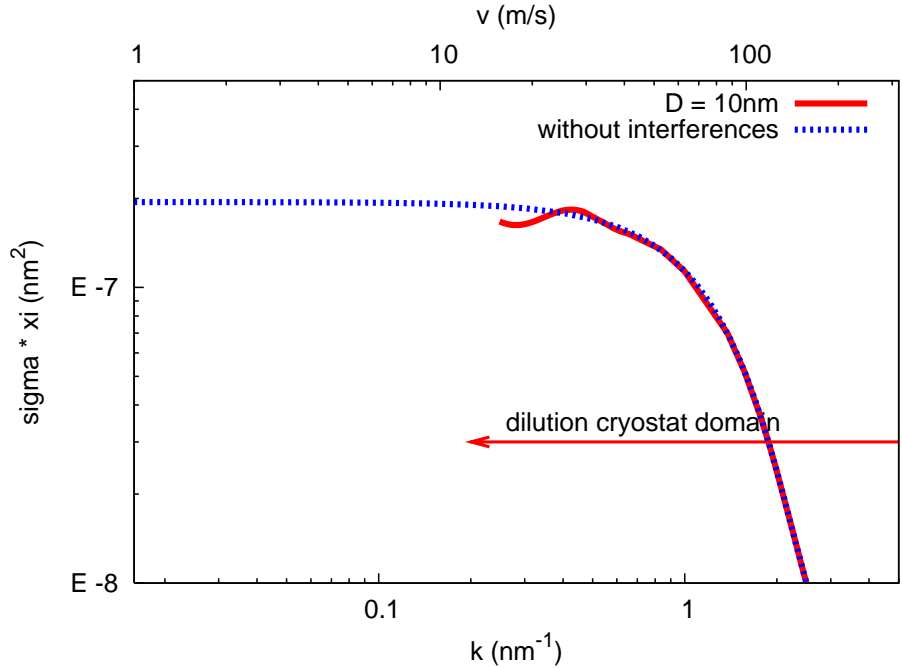}
\caption{$\xi \sigma_s$ is plotted as a function of velocity, for
a deuterium nanoparticle of radius 1 nm. The dot line is
calculated without interference effects, and the continuous line
take into account the first correction due to interference
effects.} \label{interferences}
\end{figure}

Let $\vect{D}$ be the vector distance between two nanoparticles.
At the first order of Born approximation -- neglecting multiple
diffusions -- the scattering amplitude for a neutron colliding on
this system of two nanoparticles is given by:
\begin{equation}
f(\vect{D}, \vect{q}) = \left(1+ \exp{i \vect{q} \cdot \vect{D}}\right) f(\vect{q})
\end{equation}
where $\vect{q}$ is the momentum transfer and $f(\vect{q})$ is the
scattering amplitude for the collision on a single nanoparticle.
From that result, we can conclude that the amplitude for the forward
scattering -- $\vect{q} = 0$ -- is simply twice the amplitude
calculated for a single nanoparticle. We can conclude that the
absorption cross section is not affected by the interferences.
Now, to calculate the effects of interferences on the scattering
cross section and the energy loss, the physical relevant quantity
is the average differential scattering cross section, averaging on
all possible directions for $\vect{D}$:
\begin{eqnarray}
\left\langle \deriv{\sigma_s}{\Omega} (\vect{q}) \right\rangle & = &
\int \left|f(\vect{D}, \vect{q})\right|^2 \frac{d \vect{D}/D}{4 \pi} \nonumber \\
& = & \int \left| 1+ \exp{i \vect{q} \cdot \vect{D}} \right|^2
\frac{d \vect{D}/D}{4 \pi} \left|f(\vect{q})\right|^2 \nonumber \\
& = & 2 \left(1 + \frac{\sin(q D)}{q D }\right) \left|f(\vect{q})\right|^2
\end{eqnarray}
Is is now possible to calculate $\sigma_s$ and $\xi$ using this
new effective differential cross section. Figure
\ref{interferences} shows our plot of the relevant quantity, 
$\xi \sigma_s$, we can call the energy loss cross section, and the
one calculated with the estimated correction. One can see that the
first order correction is not important.

\subsection*{Purity of Deuterium and Helium}

In all our calculations, we assumed that the helium was purely
$^4$He, with no absorption, therefore, by the Helium medium.
We also assumed that the deuterium in the nanoparticles was pure. Let us estimate the purity we actually need.

Firstly, consider the presence of hydrogen -- which is an
efficient neutron absorber -- inside the deuterium. We can compute
the purity needed so that the absorption differs from the ideal
case by less than 10 \%. If $x$ is the proportion of hydrogen
relatively to deuterium, then the absorption cross section is:
\begin{equation}
\sigma_a(x) = (1-x) \sigma_a(x=0) + x \sigma_a (x=1).
\end{equation}
Using the table \ref{nuclear} we can find that the requirement of
less than 10 \% increase is satisfied  if $x=1.5 \cdot 10^{-4}$.
We can easily achieve this purity.

Let us now consider the presence of $^3$He. We can estimate the
lifetime $\tau_a$ of a neutron due to the absorption by $^3$He. If
$x$ is the proportion of $^3$He, $\sigma_a$ is the absorption
cross section for thermal neutrons (with velocity $v_0 = 2200$
m/s), and $N_{\text{He}}$ is the density of helium nuclei, then:
\begin{equation}
\tau_a = \frac{1}{x} \frac{1}{N_{\text{He}} \sigma_a v_0}
\end{equation}
If we have a purity $x=0.5 \cdot 10^{-14}$, the absorption
lifetime $\tau_a$ is ten times the intrinsic neutron lifetime. The
natural abundance of $^3$He is $1.4 \cdot 10^{-6}$, and the
purity we need is achievable.

\section{Conclusions}

A new concept for producing high UCN density is analyzed within
the framework of the free nanoparticles model. This concept is
based on neutron cooling using ultracold nanoparticles of
deuterium, heavy water, etc. We have shown that increase in the
phase space density of neutrons, within the model of free
nanoparticles, is possible, given certain parameters of nanoparticles
and neutron velocity. Thus, solid deuterium, which is shown to
be the best material, provides efficient cooling of neutrons in
the range 1 - $10^2$ m/s, in an infinite medium of free
nanoparticles of radius 1 - 2 nm, sufficiently spatially separated. The characteristic cooling time is much shorter than the
corresponding absorption time, or the neutron $\beta$-decay
lifetime for optimum parameters of neutrons and nanoparticles. The
moderator size of, at most, a few times 10 cm allows in principle
the realization of the cooling mechanism presented.

We examine the different constraints on the model, such as scattering at individual nuclei (for too
short wavelengths), excitation of internal degrees of freedom in a
nanoparticle (such as phonons), extensions of the Born
approximation description used (the partial waves expansion), neutron
optical effects due to diffraction at several nanoparticles
simultaneously, the purity of deuterium or helium. We show that
these constraints do not change our main conclusions. We do not
consider in this present article such effects as rotation of the
nanoparticles, interaction between nanoparticles, in particular
excitation of the collective degrees of freedom for nanoparticles
in gels or any breaking of the inter-nanoparticles bounds. We do not
consider neither the influence of the non-zero temperature of the
nanoparticles on the cooling process. These phenomena are expected
to be analyzed in further publications.

\appendix
\section{\label{partwave}Partial wave expansion}
\begin{figure}[h]
\includegraphics[angle=270,width=0.9\linewidth]{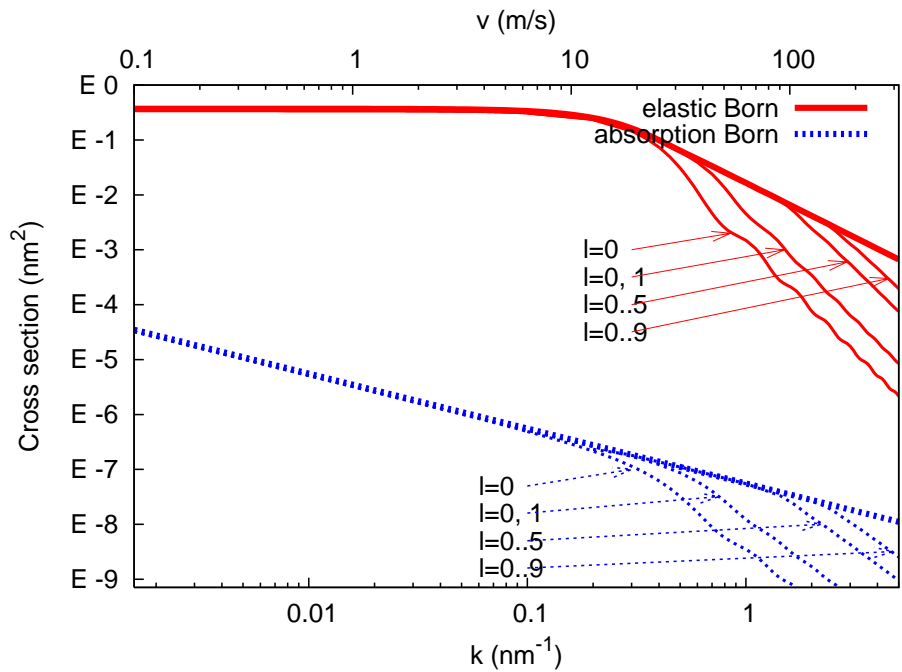}
\caption{Elastic and absorption cross-sections as a function of
neutron velocity, for a deuterium nanoparticle of radius 5 nm. The
two bold lines are calculated using the first Born approximation,
and the thin lines are calculated using the partial wave method,
till the order 0, 1, 5 and 9.} \label{partWaves}
\end{figure}

Since the potential interaction between a neutron and a
nanoparticle is approximated by spherical rectangular barrier, it
is possible to give exact solution for the scattering amplitude,
using the partial wave expansion.

The analytical solution for the scattering amplitude $\eta_l$ for
$l$th partial wave is known \cite{Taylor} to be equal to
\begin{equation}
\eta_l = \frac{\gamma_l \ h_l^-(k R) - k \ h_l^{- '} (k R))}{k \
h_l^{+ '} (k R) - \gamma_l \ h_l^+(k R)}
\end{equation}
where $h_{l}^{\pm}$ are spherical Hankel functions and
\begin{equation}
\gamma_l = K \frac{j_l'(KR)}{j_l(KR)}
\end{equation}
where $j_{l}$ is spherical Bessel function and $K^2 = k^2 - 2m(V_0
+ i V_1)/\hbar^2$.

The elastic and absorption cross sections are then given by
\begin{eqnarray}
\sigma_s  & = & \frac{\pi}{k^2} \sum_{l = 0}^{\infty}(2l+1) \left|1-\eta_l\right|^2, \\
\sigma_a  & = & \frac{\pi}{k^2} \sum_{l = 0}^{\infty} (2l + 1)
(1-\left|\eta_l\right|^2).
\end{eqnarray}

The result is shown in Fig. \ref{partWaves}, for a deuterium
nanoparticle of radius $5$ nm. We first estimated this size to be
the limit of application of the Born approximation. The figure
shows that the Born approximation is actually precise enough. The
width of the bold line in the figure corresponds to a relative
error of $5\%$, and there is no deviation between the born
approximation result and the partial wave result at this precision
level.

            %-------------------------%


\begin{thebibliography}{99}
            %-------------------------%

\bibitem{losstrap1} %OK
A.V.~Strelkov, in \emph{Proceedings of IV School on Neutron
Physics}, Alushta, Russia, 1990, p.325.

\bibitem{losstrap2} %OK
V.V.~Nesvizhevsky, A.V.~Strelkov, P.~Geltenbort, and
P.S.~Iaydjiev, Eur. J. Appl. Phys. {\bf 6}, 151 (1999).

\bibitem{losstrap3} %OK
V.V.~Nesvizhevsky, A.V.~Strelkov, P.~Geltenbort, and
P.S.~Iaydjiev, Phys. At. Nucl. {\bf 62}, 776 (1999) [Yad. Fiz.
{\bf 62}, 832 (1999)].

\bibitem{losstrap4} %OK
A.V.~Strelkov, V. V.~Nesvizhevsky, P.~Geltenbort, D.G.~Kartashov,
E.V.~Lychagin, A.Yu.~Muzychka, J.M.~Pendlebury, K.~Schreckenbach,
V.N.~Shvetsov, A.P.~Serebrov, R.R.~Taldaev, and P.S.~Iaydjiev,
Nucl. Instrum. Methods Phys. Res. {\bf A440}, 695 (2000).

\bibitem{losstrap5} %OK
L.N.~Bondarenko, P.~Geltenbort, E.I.~Korobkina, V.I.~Morozov, and
Yu.N.~Panin, Phys. At. Nucl. {\bf 65}, 11 (2002) [Yad. Fiz. {\bf
65}, 13 (2000)].

\bibitem{losstrap6} %OK
L.N.~Bondarenko, E.I.~Korobkina, V.I.~Morozov, Yu.N.~Panin,
P.~Geltenbort, and A.~Steyerl, JETP Lett. {\bf 68}, 691 (1998)
[Pis'ma Zh. Eksp. Teor. Fiz. {\bf 68}, 663 (1998)].

\bibitem{losstrap7} %OK
P.~Geltenbort, V.V.~Nesvizhevsky, D.G.~Kartashov, E.V.~Lychagin,
A.Yu.~Muzychka, G.V.~Nekhaev, V.N.~Shvetsov, A.V.~Strelkov,
A.G.~Kharitonov, A.P.~Serebrov, R.R.~Taldaev, and J.M.~Pendlebury,
JETP Lett. {\bf 70}, 170 (1999) [Pis'ma Zh. Eksp. Teor. Fiz. {\bf
70}, 175 (1999)].


\bibitem{surfacenanoparticles1} %OK
V.V.~Nesvizhevsky, Phys. At. Nucl. {\bf 65}, 400 (2002) [Yad. Fiz.
{\bf 65}, 426 (2002)].

\bibitem{surfacenanoparticles2} %OK
E.V.~Lychagin, D.G.~Kartashov, A.Yu.~Muzychka, V.V.~Nesvizhevsky,
G.V.~Nekhaev, and A.V.~Strelkov, Phys. At. Nucl. {\bf 65}, 1995
(2002) [Yad. Fiz. {\bf 65}, 2052 (2000)].

\bibitem{surfacenanoparticles3} %OK
E.V.~Lychagin, A.Yu.~Muzychka, V.V.~Nesvizhevsky, G.V.~Nekhaev,
R.R.~Taldaev, and A.V.Strelkov, Phys. At. Nucl. {\bf 63}, 548
(2000) [Yad. Fiz. {\bf 63}, 609 (2000)].

\bibitem{surfacenanoparticles4} %0K
E.V.~Lychagin, A.Yu.~Muzychka, V.V.~Nesvizhevsky, G.V.~Nekhaev,
and A.V.~Strelkov, JETP Lett. {\bf 71}, 447 (2000) [Pis'ma Zh.
Eksp. Teor. Fiz. {\bf 71}, 657 (2000)].

\bibitem{surfacenanoparticles5} %OK
V.V.~Nesvizhevsky, E.V.~Lychagin, A.Yu.~Muzychka, G.V.~Nekhaev,
and A.V.~Strelkov, Phys. Lett. {\bf B479}, 353 (2000).

\bibitem{surfacenanoparticles6} %OK
V.V.~Nesvizhevsky, Phys. Usp.  {\bf 46}, 93 (2004) [Uspekhi
Fizicheskikh Nauk {\bf 146}, 102 (2004)].

\bibitem{lifetime1} %OK
W.~Mampe, P.~Ageron, C.~Bates, J.M.~Pendlebury, A.~Steyerl, Phys.
Rev. Lett. {\bf 63}, 593 (1989).

\bibitem{lifetime2} %OK
V.V.~Nesvizhevsky, A.P.~Serebrov, R.R.~Taldaev, A.G.~Kharitonov,
V.P.~Alfimenkov, A.V.~Strelkov, and V.N.~Shvetsov, Sov. Phys. JETP
{\bf 75}, 405 (1992) [Zhurnal Eksperimental'noi i Teoreticheskoi
Fiziki {\bf 102}, 740 (1992)].

\bibitem{lifetime3} %OK
W.~Mampe, L.N.~Bondarenko, V.I.~Morozov, Yu.N.~Panin, and
A.I.~Fomin, JETP Lett. {\bf 57}, 82 (1993) [Pis'ma Zh. Eksp. Teor.
Fiz. {\bf 57}, 77 (1993)].

\bibitem{lifetime4} %OK
S.~Arzumanov, L.~Bondarenko, S.~Chernyavsky, W.~Drexel, A.~Fomin,
P.~Geltenbort, V.~Morozov, Yu.~Panin, J.~Pendlebury, and
K.~Schreckenbach, Phys. Lett. {\bf B483}, 15 (2000).

\bibitem{lifetime5} %OK
A.~Pichlmaier, J.~Butterworth, P.~Geltenbort, H.~Nagel,
V.V.~Nesvizhevsky, S.~Neumaier, K.~Schreckenbach, E.~Steichele,
and V.~Varlamov, Nucl. Instrum. Methods Phys. Res. {\bf 440A}, 517
(2000).


\bibitem{nEDM1} %OK
I.S.~Altarev, Yu.V.~Borisov, N.V.~Borovikova, S.N.~Ivanov,
E.A.~Kolomenskii, M.S.~Lasakov, V.M.~Lobashev, V.A.~Nazarenko,
A.N.~Pirozhkov, A.P.~Serebrov, Yu.A.~Sobolev, R.R.~Taldaev,
E.V.~Shulgina, and A.I.~Yegorov, Phys. Lett. {\bf B276}, 242
(1987).

\bibitem{nEDM2} %OK
P.G.~Harris, C.A.~Baker, K.~Green, P.~Iaydjiev, S.~Ivanov,
D.J.R.~May, J.M.~Pendlebury, D.~Shiers, K.F.~Smith, M.~van der
Gritten, and P.~Geltenbort, Phys. Rev. Lett. {\bf 82}, 904 (1999).


\bibitem{gravitation1} %OK
V.V.~Nesvizhevsky, H.G.~B\"orner, A.K.~Petukhov, H.~Abele,
S.~B\"assler, F.J.~Ruess, Th.~St\"oferle, A.~Westphal, A.M.~Gagarski,
G.A.~Petrov, and A.V.~Strelkov, Nature {\bf 415}, 297 (2002).

\bibitem{gravitation2} %OK
V.V.~Nesvizhevsky, H.G.~B\"orner, A.M.~Gagarski, A.K.~Petukhov,
G.A.~Petrov, H.~Abele, S.~B\"assler, G.~Divkovic, F.J.~Ruess,
Th.St\"oferle, A.~Westphal, A.V.~Strelkov, K.V.~Protasov, and
A.Yu.~Voronin, Phys. Rev. {\bf D67}, 102002 (2003).

\bibitem{gravitation3} %OK
V.V.~Nesvizhevsky, A.K.~Petukhov, H.G.~B\"orner, T.A.~Baranova,
A.M.~Gagarski, G.A.~Petrov, K.V.~Protasov, A.Yu.~Voronin,
S.~B\"assler, H.~Abele, A.~Westphal, L.~Lucovac, Eur. Phys. J. {\bf
C40}, 479 (2005).


\bibitem{charge} %OK
Yu.V.~Borisov, N.V.~Borovikova, A.V.~Vasilyev, L.A.~Grigorieva,
S.N.~Ivanov, N.T.~Kashukeev, V.V.~Nesvizhevsky, A.P.~Serebrov,
P.S.~Iadjiev, J. Tech. Phys. {\bf 58}, 1 (1988) [Zh. Tech. Fiz.
{\bf 58}, 951 (1988)].


%\bibitem{VVN1} V.V.~Nesvizhevsky, Physics of
%Atomic Nuclei, {\bf 65}, 400 (2002) [Yadernaya Fizika {\bf 65},
%426 (2002)].

\bibitem{GP} G. Pignol.
\emph{Une nouvelle source de neutrons ultra-froids : le modérateur
à nanoparticules}. Rapport de stage, Ecole Polytechnique (2005).

\bibitem{VVN2} %OK
V.V.~Nesvizhevsky, G.~Pignol, and K.V.~Protasov.
\emph{Thermalisation of neutrons by ultracold nanoparticles}. To
be published in APS proceedings of the {\sl 24th International
Conference on Low Temperature Physics}  (Orlando, Florida, USA,
10--17 August 2005).


\bibitem{othersources1} %OK
P.~Ageron, P.~De Beaucourt, H.D.~Harig, A.~Lacaze, and
M.~Livolant, Cryogenics {\bf 9}, 42 (1969).

\bibitem{othersources2} %OK
A.~Steyerl, H.~Nagel, F.-X.~Schreiber, K.-A.~Stenhauser,
R.~G\"ahler, W.~Gl\"aser, P.~Ageron, J.M.~Astruc, W.~Drexel,
G.~Gervais, and W.~Mampe, Phys. Lett. {\bf A116}, 347 (1986).

\bibitem{othersources3} %OK
R.~Golub and J.M.~Pendlebury, Phys. Lett. {\bf A62}, 337 (1977).

\bibitem{othersources4} %OK
I.S.~Altarev, N.V.~Borovikova, A.P.~Bulkin, V.V.~Vesna,
E.A.~Garusov, L.A.~Grigorieva, A.I.~Egorov, B.G.~Erozolimski,
A.N.~Erykalov, A.A.~Zakharov, S.N.~Ivanov, V.Ya.~Kezerashvili,
S.G.~Kirsanov, E.A.~Kolomenski, K.A.~Konoplev, I.A.~Kuznetsov,
V.M.~Lobashev, N.F.~Maslov, V.A.~Mitykhlyaev, I.S.~Okunev,
B.G.~Peskov, Yu.V.~Petrov, P.G.~Pikulik, N.A.~Pirozhkov,
G.D.~Porsev, A.P.~Serebrov, Yu.V.~Sobolev, R.R.~Taldaev,
V.A.~Shustov, and A.F.Shchebetov, JETP Lett. {\bf 44}, 344 (1986)
[Pis'ma Zh. Eksp. Teor. Fiz. {\bf 44}, 269 (1986)].

\bibitem{othersources5} %OK
A.P.~Serebrov, V.A.~Mityukhliaev, A.A.~Zakharov,
V.V.~Nesvizhevsky, and A.G.~Kharitonov, JETP Lett. {\bf 59}, 757
(1994) [Pis'ma Zh. Eksp. Teor. Fiz. {\bf 59}, 728 (1994)].

\bibitem{othersources6} %OK
A.~Saunders, J.M.~Anaya, T.J.~Bowles, B.W.~Filippone,
P.~Geltenbort, R.E.~Hill, M.~Hino, S.~Hoedl, G.E.~Hogan, T.M.~Ito,
K.W.~Jones, T.~Kawai, K.~Kirch, S.K.~Lamoreaux, C.Y.~Liu,
M.~Makela, L.J.~Marek, J.W.~Martin, C.L.~Morris, R.N.~Mortensen,
A.~Pichlmaier, S.J.~Seestrom, A.~Serebrov, D.~Smith, W.~Teasdale,
B.~Tipton, R.B.~Vogelaar, A.R.~Young, and J.~Yuan, Phys. Lett.
{\bf B593}, 55 (2004).


\bibitem{Gel1} %OK
L.T.~Mezhov-Deglin, A.M.~Kokotin, JETP Lett. {\bf 70}, 756 (1999)
[Pis'ma Zh. Eksp. Teor. Fiz. {\bf 70}, 744 (1999)].

\bibitem{Gel2} %OK
E.P.~Bernard, R.E.~Boltnev, V.V.~Khmelenko, V.~Kiryukhin,
S.I.~Kiselev, and D.M.~Lee, Phys. Rev. {\bf B69}, 104201 (2004).

\bibitem{Gel3} %OK
E.P.~Bernard, R.E.~Boltnev, V.V.~Khmelenko, V.~Kiryukhin,
S.I.~Kiselev, and D.M.~Lee, J. Low Temp. Phys. {\bf 134}, 169
(2004).

\bibitem{Gel4}
R.E.~Boltnev, I.N.~Krushinskaya, A.A.~Pelmenev, E.A.~Popov, D.Y.Stolyarov, and V.V.Khmelenko, J. Low Temp. Phys. {\bf 31}, 547
(2005).

\bibitem{NIST} %OK
National Institute of Standards and Technology
(www.ncnr.nist.gov/resources/n-lengths)

\bibitem{Taylor} %OK
J.R. Taylor, \emph{Scattering Theory: The Quantum Theory on
Relativistic Scattering} (John Wiey and Sons, New York, London,
Sydney, Toronto, 1972) p. 210.

\bibitem{react} %OK
S.~Glasstone, A.~Sesonske, \emph{Nuclear Reactor Engineering}, 4th
Edition, V. I., (Chapman and Hall, An International Thomson
Publishing Company, New York, Albany, Bonn, Boston, Cincinnati,
Detroit, London, Madrid, Melbourne, Mexico City, Pacific Grove,
Paris, San Francisco, Singapore, Tokyo, Toronto, Washington,
1994), p. 173.

\bibitem{Barjon} %OK
R. Barjon, \emph{Physique des réacteurs nucléaires} (Robert
Barjon, Grenoble, 1993).

\end{thebibliography}
\end{document}